\documentclass[twocolumn,secnumarabic,amssymb, nobibnotes, aps, prl]{revtex4-2}

\usepackage{color}
\usepackage[pdftex]{graphicx}
\usepackage{multirow}

\begin{document}

\title{Electronic Damage Suppression in X-ray Diffraction with Attosecond X-ray Pulses}
\author{Vladimir Lipp$^{1,2,3}$}
\email{vladimir.lipp@xfel.eu}
%\thanks{Corresponding author}
\author{Ichiro Inoue$^{4,5,6}$}
\email{ichiro.inoue@edu.k.u-tokyo.ac.jp}
%\thanks{Corresponding author}
\author{Beata Ziaja$^{3,2}$}
\affiliation{
$^1$European XFEL, Holzkoppel 4, 22869 Schenefeld.\\
$^2$Institute of Nuclear Physics, Polish Academy of Sciences, Radzikowskiego 152, 31-342 Krakow, Poland.\\
$^3$Center for Free-Electron Laser Science CFEL, Deutsches Elektronen-Synchrotron DESY, Notkestr. 85, 22607 Hamburg, Germany.\\
$^4$Department of Advanced Materials Science, The University of Tokyo, Chiba 277-8561, Japan.\\
$^5$RIKEN SPring-8 Center, 1-1-1 Kouto, Sayo 679-5148, Hyogo, Japan.\\
$^6$Institute for Experimental Physics/CFEL, University of Hamburg, Luruper Chaussee 149, 22761 Hamburg, Germany.
}
%\date{}%
\begin{abstract}
So far, the focus of state-of-the-art structure determination using x-ray free-electron laser (XFEL) pulses has been on macromolecular crystallography. This approach has achieved remarkable success in solving the structure of microcrystals smaller than a few micrometres, which are difficult to investigate using conventional synchrotron sources. However, successful applications of XFELs to inorganic and small-molecule crystallography have been limited due to rapid electron excitation in these samples. This same issue has also prevented the use of XFELs in charge-density studies of solids. 
In this study, we propose using hard x-ray pulses with an attosecond duration for x-ray imaging of charge density in organic and inorganic systems. Simulations of irradiated diamond and silicon showed that attosecond x-ray pulses consistently reduce the electronic excitation for a fixed photon energy and pulse fluence. Further reduction of transient electronic damage can be achieved by increasing the x-ray photon energy under these conditions. These theoretical predictions demonstrate a powerful and well-founded strategy for visualizing valence-electron distribution, opening up new prospects for diffraction imaging studies with XFELs.
\end{abstract}
\maketitle

Valence electrons play a key role in determining properties of solids \cite{Kittel1986}. For example, the degrees of freedom associated with valence electron orbitals are related to physical properties such as magnetism, dielectric behavior, optical characteristics, and the emergence of superconductivity. Fluctuations in the valence states of ions can lead to novel electronic states. The nature and strength of chemical bonds also strongly influence mechanical properties of condensed matter systems.

Among the available experimental techniques, x-ray diffraction uniquely enables direct visualization of valence-electron density with minimal model dependence \cite{Coppens1997, Koritsanszky2001}. Although such measurements have become routine at synchrotron facilities and are now feasible with advanced laboratory x-ray sources, their accuracy remains fundamentally limited by radiation damage, which inevitably modifies the electronic and atomic structure of the crystal during data collection.
 
The  advent of x-ray free-electron lasers (XFELs) \cite{Saldin1999, McNeilNP2010}, which generate femtosecond x-ray pulses, has opened new opportunities in crystallography. 
When an XFEL pulse interacts with a sample, photoelectrons are instantaneously emitted. After atom-specific recombination times have elapsed, Auger electrons are released as core holes are filled. The subsequent collisional ionizations triggered by photo- and Auger electrons proceed on the femtosecond time scale \cite{Ziaja2001, Ziaja2005}. 
These electronic excitations modify the 
atomic potential energy surface and ultimately lead to atomic disorder.
However, numerical studies have shown that when the x-ray pulse is shorter than the onset of atomic motion, diffraction data can be collected before significant structural damage occurs \cite{Neutze2000, Chapman2014, Ilme2015}, even at absorbed doses far exceeding the conventional dose limit for crystallography \cite{Owen2006, Howells2009}.

To date, structure determination using XFEL pulses has been primarily focused on macromolecular crystallography. Applications of XFELs to inorganic and small-molecule crystallography, which are the main targets of conventional x-ray diffraction, remained limited, with only a few reported examples \cite{Schriber2022, Takaba2023}.
Although these studies have enabled the determination of atomic structures in radiation-sensitive materials, direct visualization of valence-electron density remains challenging.
A major obstacle is the ultrafast x-ray-induced electronic excitation, which has been observed in recent experiments
For example, femtosecond x-ray pump--x-ray probe experiments on diamond \cite{Inoue2021} and silicon \cite{Inoue2025} have shown that valence electrons are strongly excited within a few femtoseconds. Furthermore, diffraction measurements on silicon \cite{Inoue2023} revealed that not only valence electrons but also inner-shell electrons are excited when using nanofocused XFEL pulses.
Because these electronic excitations occur on a femtosecond timescale, diffraction measurements with XFELs cannot directly probe the pristine electron-density distribution.

In this Letter, we investigate the potential of attosecond x-ray pulses to suppress radiation-induced electronic excitation. Attosecond x-ray pulses generated by highly compressed electron beams have recently become available at facilities such as LCLS \cite{MarinelliAPL2017, HuangPRL2017, Inoue2025_arxiv} and European XFEL \cite{Yan2024}. Our numerical simulations demonstrate that these emerging x-ray sources provide a pathway toward x-ray diffraction experiments free from both electronic excitation and structural damage.

%%%%%%%%%%%%%%%%%%%%%%%%%%%%%%%%%%%%%%%%%%%%%%%%
% Diamond, 6 keV
\begin{figure*}[htbp]
		\includegraphics[width=11cm]{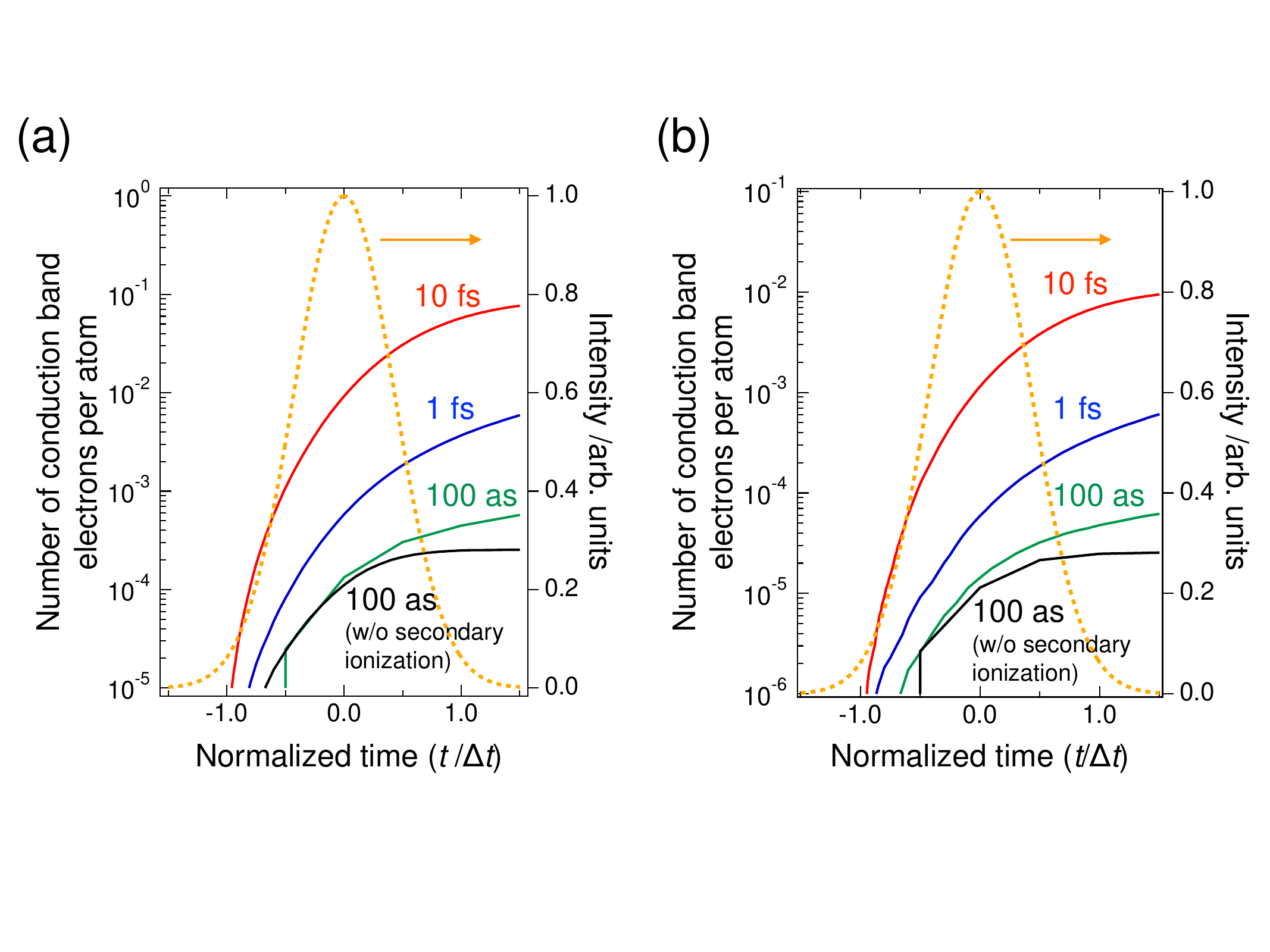}
		\caption{Number of excited (conduction band) electrons per atom plotted as a function of time in diamond irradiated with 6-keV x-ray pulses of the following fluences: (a) 10 $\mu$J/$\mu$m$^2$ or (b) 1 $\mu$J/$\mu$m$^2$. The horizontal axis shows the evolution time divided by the x-ray pulse duration, $t/\Delta t$, here equal to 10 fs (red), 1 fs (blue), or 100 as (green). Black curves represent the results from simulations for 100-as pulses, with electron impact ionization not taken into account. Orange dotted lines show the intensity envelope of the pulses (in arbitrary units).}
\label{fig:diamond}
\end{figure*}

%%%%%%%%%%%%%%%%%%%%%%%%%%%%%%%%%%%%%%%%%%%%%%%%%%%
% Silicon, 6 keV
\begin{figure*}[htbp]
		\includegraphics[width=11.5cm]{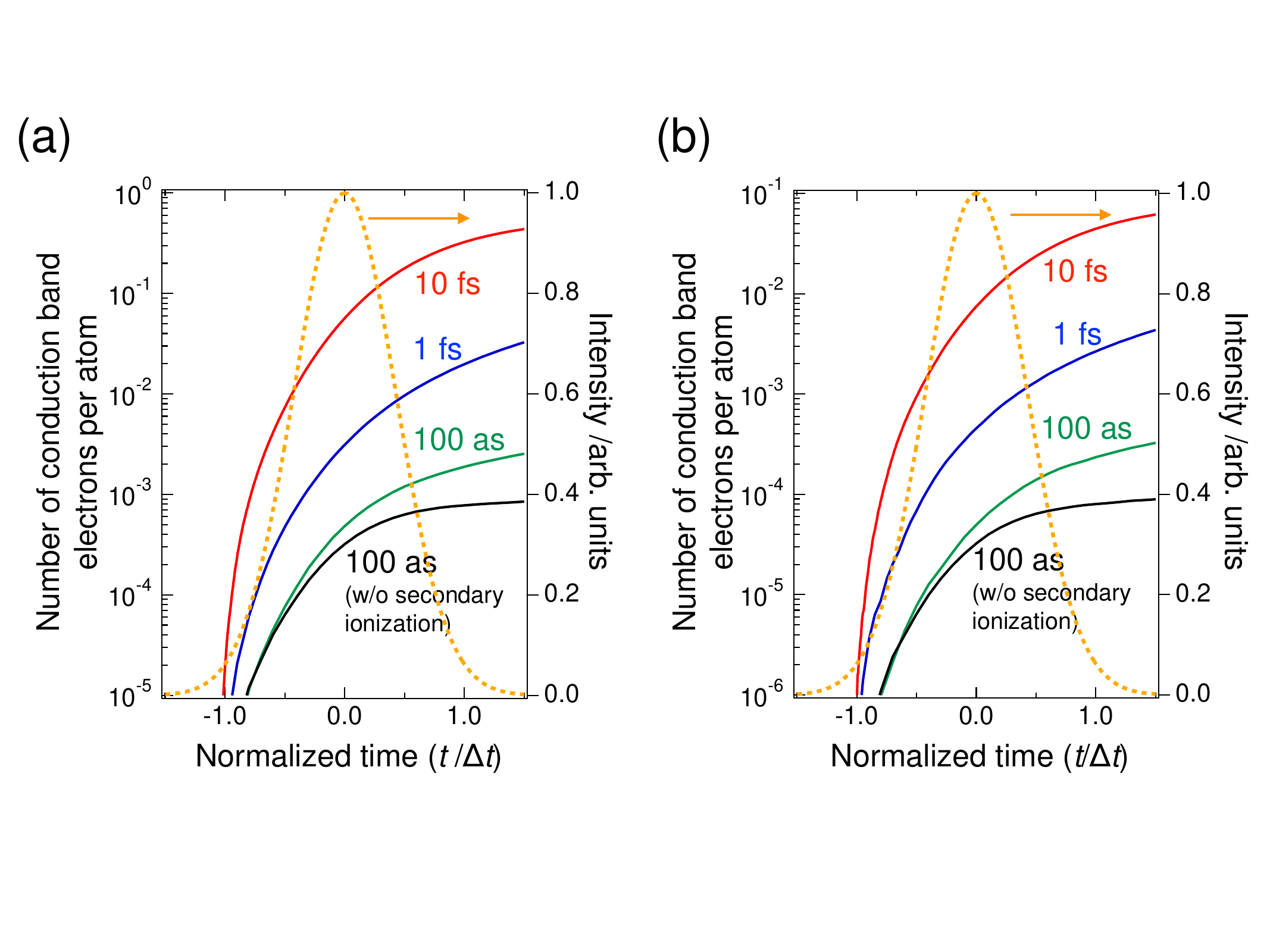}
		\caption{Number of excited (conduction band) electrons per atom plotted as a function of time in silicon irradiated with 6-keV x-ray pulses of the following fluences: (a) 10 $\mu$J/$\mu$m$^2$ or (b) 1 $\mu$J/$\mu$m$^2$. The horizontal axis shows the evolution time divided by the x-ray pulse duration, $t/\Delta t$, here equal to 10 fs (red), 1 fs (blue), or 100 as (green). Black curves represent the results from simulations for 100-as pulses, with electron impact ionization not taken into account. Orange dotted lines show the intensity envelope of the pulses (in arbitrary units).}
\label{fig:silicon}
\end{figure*}

%%%%%%%%%%%%%%%%%%%%%%%%%%%%%%%%%%%%%%%%%%%%%%%%%%

\begin{figure*}[htbp]
		\includegraphics[width=11.5cm]{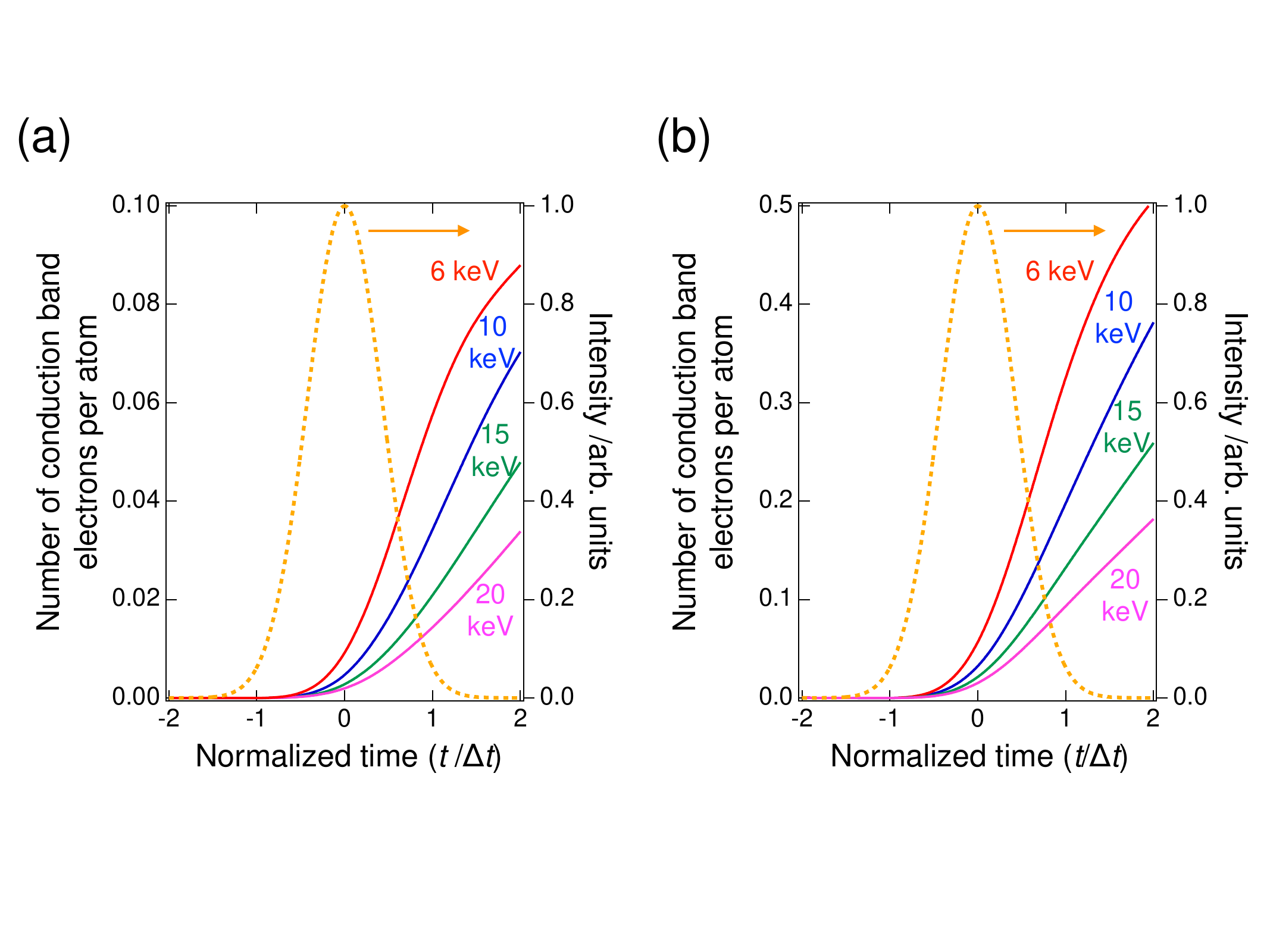}
		\caption{Number of excited (conduction band) electrons per atom plotted as a function of time for (a) diamond and (b) silicon irradiated with 10-fs pulses and photon energies of 6, 10, 15, and 20 keV at the following fixed absorbed doses: (a) 1.35 eV/atom, (b) 4.12 eV/atom. The horizontal axis shows the evolution time divided by the x-ray pulse duration, $t/\Delta t$, here equal to 10 fs (red), 1 fs (blue), or 100 as (green). Black curves represent the results from simulations for 100-as pulses, with electron impact ionization not taken into account. Orange dotted lines show the intensity envelope of the pulses (in arbitrary units). }
\label{fig:photon-energy}
\end{figure*}

In what follows, we will investigate electron excitation in diamond and silicon. The electronic structure of these materials is relatively simple, allowing us to simulate electronic excitation with high precision. Diamond can be regarded as a representative model system for investigating the feasibility of electron-density determination in organic molecules. Similarly, silicon represents a model system for  analysis of  electronic damage in inorganic materials.

We have used our in-house code XTANT+ (x-ray-induced Thermal and Nonthermal Transitions Plus) \cite{lipp2022density} to perform numerical simulations of ultrafast x-ray-induced electron excitations. This hybrid code is based on a combination of Monte Carlo, density-functional tight binding (DFTB) and molecular dynamics simulation approaches. Introduced in 2022 \cite{lipp2022density}, XTANT+ has already provided insights into x-ray-induced graphitization of diamond \cite{lipp2022density,Heimann2023Nonthermal} and x-ray-induced atomic displacements in corundum on femtosecond timescales \cite{Inoue2022Delayed}. In this work, we have extended the software to simulate x-ray irradiated silicon based on a DFTB parametrization from Ref \cite{Sieck2000Structure}. 

All simulations were performed for a 512-atom-large supercell (4$\times$4$\times$4 unit cells) irradiated with a spatially uniform x-ray pulse of various photon energies, FWHM (full width at half maximum) pulse durations, and absorbed doses. Periodic boundary conditions were applied. The Monte-Carlo module, simulating photoabsorption, Auger decay and collisional ionization of atoms by high-energy electrons (in the present simulations, the electrons with kinetic energy above 10 eV), was set to perform 10$^5$ Monte-Carlo iterations for a reliable statistics. The DFTB module based on the well-known DFTB+ software \cite{dftbp2020,dftbplus2025} calculated transient band structure and interatomic forces at each time step. The electrons with energies below 10 eV were assumed to obey the Fermi distribution on the DFTB-calculated transient electronic levels. The calculated DFTB interatomic forces were used by the molecular dynamics module to predict atomic trajectories. The duration of a single time step was set to 10 as, which ensured reliable conservation of the total energy. More details about the employed computational scheme can be found in Ref. \cite{lipp2022density}. 

In addition to simulations that included all relevant electron excitation processes, we also performed simulations that included only photoinduced processes (photoionization and Auger decay). This mimicked the case in which most energetic electrons quickly escaped from the beam focus. Accordingly, we then switched off all electronic collisional ionizations of atoms by setting the corresponding mean free paths to infinity.

First, we discuss the evolution of the number of excited electrons depending on x-ray pulse duration. Figures \ref{fig:diamond}a and \ref{fig:diamond}b show the number of excited (conduction-band) electrons per atom in diamond irradiated with 6-keV pulses of three different pulse durations (FWHM): $\Delta t = 10$ fs, 1 fs, and 0.1 fs. The horizontal axis shows the evolution time divided by the pulse duration.

For each pulse duration, two fluence  values were considered: 10 $\mu$J/$\mu$m$^2$ (Fig. \ref{fig:diamond}a) and 1 $\mu$J/$\mu$m$^2$ (Fig. \ref{fig:diamond}b). In case of the photon energy of 6 keV, the previous value corresponds to the absorbed dose of 1.35 eV/atom, which is close to the ultrafast structural damage threshold (about 1.5 eV/atom \cite{lipp2022density}) and a typical value used in the current SFX experiments using microfocused XFEL pulses. For comparison, the results obtained by including only photoabsorption and Auger decay (i.e., excluding secondary ionization processes) are shown with black curves.

For both fluence values,  the electron excitation during the pulse was suppressed for shorter pulses. These results confirm that femtosecond pulses are not sufficiently short to suppress secondary electron excitation within the pulse duration. The simulation results obtained for 100 as pulses with and without secondary ionization are very similar, indicating that sub-femtosecond  pulses effectively outrun nearly all secondary electron generation processes during the pulse.

The outrunning of secondary electron excitation can also be observed in silicon. Figures \ref{fig:silicon}a and \ref{fig:silicon}b show the number of conduction-band electrons per atom in silicon irradiated with 6-keV pulses with $\Delta t =$ 10 fs, 1 fs, and 0.1 fs, for the same set of pulse fluences.  As in the case of diamond, shorter pulses lead to a pronounced suppression of secondary electron excitation during the pulse duration. For 100 as pulses, the results obtained with and without secondary ionization are again close to each other, confirming that sub-femtosecond pulses effectively outrun secondary electron cascading.

Because the measured  diffraction signal is time integrated, i.e., it is time integral of the instantaneous scattering signal weighted by the temporal pulse intensity envelope, a pulse-intensity-weighted average is a natural measure of the expected probe signal strength. Recent XFEL damage studies on silicon analyzed similar quantities such as diffraction efficiency \cite{inoue2023femtosecond,ziaja2023modeling}, whereas a recent study on electron dynamics in silicon explicitly introduced pulse-weighted silicon ionization degree \cite{cardoch2025modeling}. A related work on single-particle-imaging has introduced pulse-weighted average charge \cite{banerjee2026impact}. Motivated by this logic, to quantitatively assess the feasibility of an electronically damage-free diffraction experiment, we define the pulse-intensity-averaged number of conduction-band electrons per atom:
%%%%%%%%%%%%%%%%
\begin{equation}
   \bar{n}_c=\frac{\int_{-\infty}^{\infty} I(t) n_c(t) dt}{\int_{-\infty}^{\infty} I(t) dt},
\end{equation}
%%%%%%%%%%%%%%%%
where $t$ is the time, $I$ is the instantaneous intensity of x-ray pulse, $n_c$ is the number of conduction band electrons per atom.

This quantity can serve as a measure of the influence of electronic damage on the observed diffraction signal.

Table \ref{table:diamond} shows the values of $\bar{n}_c$ in x-ray irradiated diamond at the fluence of 10 $\mu$J/$\mu$m$^2$, for different pulse durations (10 fs, 1 fs, and 100 as) and photon energies (6 keV, 10 keV, 15 keV, and 20 keV).

%%%%%%%%%%%%%%%%%%%

\begin{table}[htbp]
\centering
{\small
\caption{Pulse-intensity-averaged number of conduction-band electrons per atom in x-ray irradiated diamond at the pulse fluence of 10 $\mu$J/$\mu$m$^2$.}
\label{table:diamond}
\begin{tabular}{c c c c}
\hline
&\multicolumn{3}{c}{Pulse duration}\\
\cline{2-4}
Photon energy &\multicolumn{1}{c}{10 fs}&\multicolumn{1}{c}{1 fs}
&\multicolumn{1}{c}{100 as}\\
\hline
6 keV & \multicolumn{1}{c}{1.36$\times$10$^{-2}$} & \multicolumn{1}{c}{8.38$\times$10$^{-4}$} & \multicolumn{1}{c}{4.25$\times$10$^{-4}$}  \\
10 keV & \multicolumn{1}{c}{7.26$\times$10$^{-3}$} & \multicolumn{1}{c}{4.34$\times$10$^{-4}$} & \multicolumn{1}{c}{1.05$\times$10$^{-4}$}  \\
15 keV & \multicolumn{1}{c}{4.35$\times$10$^{-3}$} & \multicolumn{1}{c}{2.61$\times$10$^{-4}$} & \multicolumn{1}{c}{6.49$\times$10$^{-5}$}  \\
20 keV & \multicolumn{1}{c}{3.01$\times$10$^{-3}$} & \multicolumn{1}{c}{1.86$\times$10$^{-4}$} & \multicolumn{1}{c}{3.66$\times$10$^{-5}$} \\
\hline
\end{tabular}
}
\end{table}

Given that a typical reliability factor in charge-density analysis by x-ray crystallography is on the order of a few percent (e.g., \cite{ptasiewicz1999charge,Chuang2017Charge,Krause2017Validation}), it is desirable that the intensity-weighted number of conduction band electrons per atom should be less than 10$^{-2}$ \cite{Tanaka2025Charge,nakashima2017quantitative}. For all combinations of pulse duration and photon energy shown in Table \ref{table:diamond}, except for a photon energy of 6 keV with a pulse duration of 10 fs, the value of  $\bar{n}_c$ satisfies this condition. This indicates that visualization of electron-density distributions in organic materials is feasible under typical photon parameters used in current SFX experiments (fluence of 10 $\mu$J/$\mu$m$^2$, pulse duration of $\sim$10 fs).

Recent developments in nanofocusing optics at XFEL facilities have enabled the generation of tightly focused beams with fluences up to two orders of magnitude higher compared with the beam used for SFX experiments \cite{Mimura2014, Nagler2017, YumotoAS2020, Yamada2024}. The application of such high-intensity pulses will open new opportunities for imaging of valence electrons in small molecular samples. However, the increase in x-ray fluence inevitably enhances electronic excitation during the pulse. Assuming that the number of conduction-band electrons per atom is proportional to the fluence under the pulse-duration conditions listed in Table \ref{table:diamond}, $\bar{n}_c$ exceeds $10^{-2}$ for femtosecond pulses. In contrast, $\bar{n}_c$ still remains less than $10^{-2}$ for attosecond pulses, indicating that pristine electrons can be visualized in organic materials using attosecond x-ray pulses.

Table \ref{table:silicon} shows the values of $\bar{n}_c$ in x-ray irradiated silicon at the fluence of 10 $\mu$J/$\mu$m$^2$, which is also a typical value in the current SFX experiments. For 10 fs pulses, $\bar{n}_c$ exceeds $10^{-2}$, indicating that significant electronic damage in inorganic materials occurs under these conditions. To visualize the valence-electron density distributions in their pristine state, ultrashort pulses with durations on the order of $\sim$1 fs are required. The use of attosecond pulses further reduces $\bar{n}_c$, thereby allowing an increase in fluence by up to an order of magnitude than microfocused XFEL pulses while still enabling visualization of pristine electron-density distributions. 

%%%%%%%%%%%%%%%%%%%%%%%%%%%%%%%%%%%%%%%
\begin{table}[htbp]
\centering
{\small
\caption{Pulse-intensity-averaged number of conduction-band electrons per atom in x-ray irradiated silicon at the pulse fluence of 10 $\mu$J/$\mu$m$^2$.}
\label{table:silicon}
\begin{tabular}{c c c c}
\hline
&\multicolumn{3}{c}{Pulse duration}\\
\cline{2-4}
Photon energy &\multicolumn{1}{c}{10 fs}&\multicolumn{1}{c}{1 fs}
&\multicolumn{1}{c}{100 as}\\
\hline
6 keV & \multicolumn{1}{c}{5.72$\times$10$^{-2}$} & \multicolumn{1}{c}{4.46$\times$10$^{-3}$} & \multicolumn{1}{c}{1.28$\times$10$^{-3}$}  \\
10 keV & \multicolumn{1}{c}{3.26$\times$10$^{-2}$} & \multicolumn{1}{c}{2.40$\times$10$^{-3}$} & \multicolumn{1}{c}{7.08$\times$10$^{-4}$}  \\
15 keV & \multicolumn{1}{c}{2.16$\times$10$^{-2}$} & \multicolumn{1}{c}{1.60$\times$10$^{-3}$} & \multicolumn{1}{c}{4.53$\times$10$^{-4}$}  \\
20 keV & \multicolumn{1}{c}{1.53$\times$10$^{-2}$} & \multicolumn{1}{c}{1.19$\times$10$^{-3}$} & \multicolumn{1}{c}{3.42$\times$10$^{-4}$} \\
\hline
\end{tabular}
}
\end{table}
%%%%%%%%%%%%%%%%%%%%%%%%%%%%%%%%%%%%%%

Next, we examine the dependence of electron excitation on photon energy (Fig. \ref{fig:photon-energy}). To identify the optimal photon energy for suppressing electronic excitation during pulse irradiation, we compare simulation results on the number of excited (conduction band) electrons for the two materials at four different photon energies (6, 10, 15, and 20 keV). We find that electron excitation decreases with increasing photon energy.

As an example, Figures \ref{fig:photon-energy}a and \ref{fig:photon-energy}b present simulations of diamond and silicon irradiated with 10-fs pulses at the absorbed doses of 1.35 eV/atom and 4.12 eV/atom, respectively, corresponding to the fluence value of 10 $\mu$J/$\mu$m$^2$ at 6 keV. Although higher photon energies are advantageous for reducing electron excitation during pulse irradiation, the achievable reduction is limited by a factor of ten at most within the photon energy range currently available at XFEL facilities (up to around 20 keV), see Tables 1, 2. Therefore, the attosecond pulses are much more efficient to suppress the electronic damage during pulse irradiation. 

In summary, we investigated the potential of attosecond x-ray pulses to suppress transient electronic excitation in charge-density studies.
Through systematic simulations of diamond and silicon, representative organic and inorganic materials, respectively, we demonstrated that shortening the x-ray pulse duration to the attosecond regime consistently reduces electronic damage at fixed photon energy and fluence.
In silicon, femtosecond pulses readily exceed the critical excitation level of 10$^{-2}$ excited electrons per atom, whereas attosecond pulses with experimentally relevant x-ray parameters maintain electronic excitation below this threshold.
Moreover, increasing the photon energy further suppresses transient electronic damage under fixed fluence and pulse duration conditions. These results establish the combination of attosecond pulse duration and high photon energy as a powerful route toward damage-free visualization of valence-electron density in solids.

\acknowledgments{This research was conducted using the computational resources of the Maxwell Cluster at Deutsches Elektronen-Synchrotron (DESY), Hamburg, Germany. The project received funding from the Polish Ministry of Science and Higher Education (2022/WK/13), KAKENHI (22KK0233, 24K21199) and the Japan Science and Technology Agency (JPMJPR24J1).}

%I.I. suggested the idea. V.L. performed the simulations with valuable input from B.Z. V.L., I.I., and B.Z. contributed to the interpretation and discussions of the results. I.I. prepared the figures and wrote the manuscript with contributions from all authors. B.Z. supervised the project.

\paragraph*{Data availability}{XTANT+ software package was used to obtain the simulated data in this article. The software package is not publicly available due to licensing conditions, but we plan to release it as free and open-source software in the future.}

%\bibstyle{natbib}
\bibliography{Ref,lipp-own,mybib}

\end{document}